# Predicted Realization of Cubic Dirac Fermion in Quasi-One-Dimensional Transition-Metal Mono-Chalcogenides

Qihang Liu[1,*] and Alex Zunger[1]

[1]Renewable and Sustainable Energy Institute, University of Colorado, Boulder, CO 80309, USA

*Email: qihang.liu85@gmail.com

## Abstract

We show that the previously predicted Fermion particle that has no analogue in the standard model of particle theory — the "cubically dispersed Dirac semimetal" (CDSM) — is realized in a specific, stable solid state system that has been made years ago, but was not appreciated to host such a unique Fermion, composed of six Weyl Fermions, 3 with left-handed and 3 with right-handed chirality. We identified the crystal symmetry constraints and found the space group $P6_3/m$ as one of the two that can support a CDSM, of which the characteristic band crossing has linear dispersion along the principle axis but cubic dispersion in the plane perpendicular to it. We then conducted a material search using density functional theory identifying a group of quasi-one-dimensional molybdenum mono-chalcogenide compounds $A^I(MoX^{VI})_3$ ($A^I$ = Na, K, Rb, In, Tl; $X^{VI}$ = S, Se, Te) as ideal CDSM candidates. Studying the stability of the $A(MoX)_3$ family reveals a few candidates such as $Rb(MoTe)_3$ and $Tl(MoTe)_3$ that are predicted to be resilient to Peierls distortion, thus retaining the metallic character. The combination of one-dimensionality and metallic nature in this family provides a platform for unusual optical signature – polarization dependent metallic vs insulating response.

**Introduction**

*The topological significance of band crossings in solids:* The crossing on energy bands in complex materials showing dense manifold of states is a ubiquitous effect routinely reported in the past ~ 50 years in countless publications, a visual effect often referred to as "band spaghetti". Such crossings have been known for a long time to result from specific space group symmetries [1,2], involving various band dispersion $E(\boldsymbol{k})$ with powers $k^n$ and to be characterized by a band degeneracy $g(\boldsymbol{k})$ at the crossing wavevector $\boldsymbol{k}$. With the renewed interest in materials with strong spin-orbit coupling (SOC) it became recently clear that such band crossing points could carry interesting information on the topological behavior of the system, leading to specific behaviors of surface/edge states in reduced dimensions [3-5]. In SOC system respecting time-reversal symmetry, the degeneracy of multi-band crossings in momentum space could be $g(\boldsymbol{k}) = 2, 3, 4, 6$, and 8, and the dispersion power $k^n$ with $n = 1, 2, 3$ at the crossing point could be linear, quadratic or cubic, respectively. Crossings between conduction and valence band with linear dispersion in at least one $k$-space direction and $g = 2$-fold degeneracy corresponds to Weyl semimetal (WSM) [3,6-10] and $g = 4$-fold degeneracy corresponds to Dirac semimetal (DSM) [4,11-16]. The remaining degeneracies at band crossing points $g = 3, 6$, and 8 correspond to quasi-particles without analogous states in the standard model of particle physics [5,17-19]. The latter respects Poincare symmetry and has but three Fermions types – Weyl, Dirac and Majorana.

*Expected special features of unconventional Dirac Fermions:* In general, Dirac/Weyl band crossings in bulk solids could be accompanied by different dispersion powers in different directions, as constrained by their crystal symmetries**.** At these band crossings (termed "Dirac/Weyl points"), the dispersion along the principle rotation axis ($c$) is linear, whereas the dispersion along the $a$-$b$ plane could be either linear ($n = 1$), quadratic ($n = 2$), or cubic ($n = 3$). A <u>Weyl point</u> with in-plane dispersion power $n$ carries a Chern number $n$ or $-n$, corresponding to a degeneracy of $n$ conventional Weyl Fermions (left- or right-handed) all with the same chirality [7]. In contrast, a <u>Dirac point</u> with specific $n$, has zero net Chern number, and is $2n$-fold degenerate Weyl Fermion with half left-handed and half right-handed chirality [15,20,21]. Such Weyl and Dirac Fermions with high-order dispersions, caused by crystalline symmetry in solids, also do not have counterparts in high energy physics. While the surface states of DSM are not topologically protected (unlike those of TI and WSM), such quadratic ($n = 2$) and *cubic* Dirac Fermions ($n = 3$), especially the latter, are interesting relative to the conventional $n = 1$ Dirac Fermions: Their distinguishing features (as discussed in the concluding remarks) include creation of special WSM with multiple Fermi-arcs; characteristic quantum transport signatures; quantum criticality and phase transitions.

***Search for materials that realize cubic Dirac Fermions:*** In addition to the efforts to define and classify such specific "new Fermion" band crossings induced by crystalline symmetries and topology in condensed matter systems [5], an important challenge is to systematically identify material realization of such unusual Fermions. We summarize all the different types of Fermions in solid state physics, classified by the degree of band degeneracy ($g$) and the highest power of band dispersion ($n$), with example materials in Appendix A. Whereas *quadratic dispersion* has been predicted to exist in $SrSi_2$ [22] ($g = 2$), band-inverted $\alpha-Sn$ ($g = 4$), and $PdSb_2$ ($g = 6$) [5], the *cubic dispersion*, expected to exist both in Weyl [7] and Dirac semimetals [15], has not been realized as yet in any material candidates. In part, the difficulty to find such cubically-dispersed Dirac semimetals (CDSM) is related to the multitude of nontrivial requirements one needs to impose on such a material search, including appropriate crystal symmetry, angular momentum and electron filling. Finding such materials by accidental discovery or simple trial-and-error would thus be unlikely.

Here we establish understanding-based design principles for CDSM and use these to deliberately screen the candidates that satisfy such conditions by exhaustively looking through all 230 space groups in 3D. We find that only materials in two space groups $P6_3/m$ (No. 176) and $P6/mcc$ (No. 192) have the potential to host cubic Dirac Fermions. This narrowing down of the possibilities is then followed up by a (yet non-exhaustive) material search of compounds belonging to these two CDSM-hosting space groups using density functional theory (DFT, see Appendix B for computational details). We identify a group of molybdenum transition-metal chalcogenide compounds $A^I (MoX^{VI})_3$ ($A^I = Na$, K, Rb, In, Tl; $X^{VI} = S$, Se, Te) with space group $P6_3/m$ as ideal candidates. The structure of this type of compound (Fig. 1) is basically quasi-1D chains $(Mo_3X_3)^{1-}$ running in the direction of the c axis and separated by the hexagonal $A^{1+}$ framework. Most of these compounds have been synthesized in 1980's [23], and preliminary electronic structure with linear dispersion along the chain direction has been reported previously [24,25]. Recently, Gibson et al. theoretically proposed this family of compounds as the conventional linearly-dispersed DSM [26]. In the present paper, we find that the A(MoX)$_3$ family exhibits (i) one cubically-dispersed Dirac point (DP) and (ii) three linearly-dispersed DPs induced by the non-symmophic symmetry within the Brillouin zone (BZ), like those envisioned for BiO$_2$ in a hypothetical unstable structure [11]; (iii) quadratic DPs and another type of linear DP along the chain direction in the conduction bands. Furthermore, we predict that because of the coexistence of reduced 1D dimensionality and metallic bands embedded in a semiconductor bulk, A(MoX)$_3$ CDSMs will exhibit polarization-dependent optical response.

**Stability under Peierls distortion:** To predict realistic CDSM candidates we systematically studied the stability of these quasi-1D A(MoX)$_3$ compounds under Peierls distortion. DFT results show that some compounds with strong 1D character inevitably experience Peierls distortion, indicating a metal-to-insulator transition below a critical temperature, consistent with the previous transport measurements [27,28]. In contrast, we find that the ground states of some compounds, e.g., Rb(MoTe)$_3$ and Tl(MoTe)$_3$, are immune to Peierls distortion even at low temperature, leading to the retention of the CDSM phase. Another stability issue that will be explored in the future is the resilience of the structure to spontaneous formation of intrinsic defects [such as A vacancies in A(MoX)$_3$] that may shift the Fermi level of the insulating phase.

**Results**

**Design principle for identifying CDSM:** This involves both crystal symmetry as well as electronic occupancy conditions. We will first consider the crystal symmetry requirement for DSM with cubic in-plane dispersions.

Here we focus on non-magnetic materials with inversion symmetry $P$ and SOC, which respects time-reversal symmetry $T$ with $T^2 = -1$. The cubic dispersion at the DP can only appear as the form of four-fold degeneracy occurring in the time-reversal invariant (TRI) k-point, whose little group has C$_6$ symmetry [15]. Given that $P$ and $T$ ensure that all two-fold degenerate bands with the two components related each other by $PT$, i.e., $\psi(\boldsymbol{k}, \boldsymbol{\sigma})$ and $PT\psi(\boldsymbol{k}, \boldsymbol{\sigma}) = \psi(\boldsymbol{k}, -\boldsymbol{\sigma})$, we need an extra pair of state $L\psi$ and $PTL\psi$ that differ from with $\psi$ and $PT\psi$, where L is an Hermitian symmetry operator that commutes with the Hamiltonian $H$. This can be achieved by finding another Hermitian symmetry operator $\mathcal{A}$ to fulfill $\{A, A_{PT}\} \cap \{A_L, A_{LPT}\} = \emptyset$, where $A$ is the eigenvalue of $\psi$ under $\mathcal{A}$. For the least symmetry required by the system we first let $L = P$, then the extra degeneracy immediately requires an anti-commutation $\{\mathcal{A}, \mathrm{P}\} = 0$, implying that $\mathcal{A}$ contains a *non-symmophic symmetry*.

Therefore, to establish cubic in-plane dispersion at $g = 4$ band crossing we need at least three symmetry filters: inversion, $C_6$ rotation and presence of non-symmophic operations such as screw axis or glide reflection. These requirements already exclude most of the space groups and leave only four possibilities, P6$_3$/m (No. 176), P6/mcc (No. 192), P6$_3$/mcm (No. 193) and P6$_3$/mmc (No. 194). All of these space groups have 4-fold degeneracy at TRI $k$-points within $k_z = \pi$ plane, forced by screw axis (No. 176, 193 and 194) or glide reflection symmetry (No. 192). However, space groups No. 193 and 194 have three mirror planes parallel to $C_6$ axis, posing extra symmetry conditions that force three high-symmetry lines to be degenerate and thus no band splitting along these directions [29]. Therefore, such nodes form an infinite network through the BZ, leading to new types of topological semimetal named nodal-line [30-35] and nodal-ring semimetal [29,36] that are distinct from DSM and WSM. Therefore, only materials with space group P6$_3$/m and P6/mcc can host cubic Dirac Fermions in terms of crystal symmetry (see Appendix C for details).

To further ensure that the leading order of the in-plane dispersion is cubic the states of conduction band and valence band at the DP should have opposite eigenvalues of $C_6$ operator [15] $i$, or -$i$. This means that the four degenerate states at the DP should have the orbital part with angular momentum $l_z = \pm 1$, e.g., $p_x \pm i p_y$ or $d_{xz} \pm i d_{yz}$ components. In addition, the 4-fold degenerate DP bunches conduction and valence band together (half filling), which requires the total number of electrons $N_e$ mod 4 = 2. On the other hand, because of non-symmophic symmetry the materials should have at least two sublattices, indicating an odd number count per formula unit (f.u.).

We summarize all the requirements for CDSM as follows: (i) inversion, (ii) $C_6$ rotation, (iii) non-symmophic symmetry, (iv) $l_z = \pm 1$ states and (v) odd number of electrons per

f.u.. Based on these design principles, we performed a design-principle guided material search and found that a group of A(MoX)₃ compounds with space group P6₃/m are ideal candidates for CDSM.

***Crystal structure and chemical bonding of the A(MoX)₃ family:*** The quasi-1D structures A(MoX)₃ system is derived from the general family, known as "Chevrel clusters" [37], that has been synthesized with extended MoX clusters $Mo_6X_8$. The $Mo_6X_8$ unit can be viewed as a $Mo_6$ octahedral surrounded by eight X chalcogen atoms, or two $Mo_3X_3$ star-shaped planes capped by two other X atoms along c axis ($C_6$ rotation axis), as shown in Fig. 1a. This unit cluster can be expanded along c axis infinitely and form an 1D chain by repeating the $Mo_3X_3$ unit along c axis, as shown in Fig. 1b and c. Such an extension provides a transition between a "molecule" to a chain-like structure, with each equidistant unit of the chain forming an equilateral triangle by three Mo atoms. The monovalent cation element A (alkali or In, Tl) forms linear chains between the $Mo_3X_3$ units, as shown in Fig. 1e.

The intra-triangle (within a-b plane) Mo-Mo bond lengths are 2.64-2.65 Å for all the A(MoX)3 compounds considered here, while the inter-triangle Mo-Mo bond ranges from 2.69-2.76 Å. These bond lengths are all well below twice of the atomic radius of Mo (2.01 Å), indicating strong Mo-Mo interaction. In addition to the ordinary ionic/valent bonds (e.g., Mo-X), the existence of Mo-Mo interaction leads to delocalized electrons along the chain and complicates the overall bonding types of Mo atoms. As a result, the average valence state of Mo can be a fractional number, offering possibilities to have an odd number of electrons per f.u. Considering A(MoX)₃ with alkali atom A($s^1$), Mo($d^5s^1$) and X ($s^2p^4$) and only Mo atom can be multi-valent, the valence state on Mo atom is thus +1.67. Since the total electrons per f.u. is an odd number 37.

In addition to the odd number of electron filling, we have identified this family of compound as a CDSM (Table I) also because of its reported space group being non-symmophic P6₃/m (No. 176) [23]. The structure contains 12 symmetry operations: identity, inversion, screw axis operation $\{C_6|(0,0,1/2)\}$, which is a six-fold rotation about c axis followed by a fractional lattice translation c/2, and their combinations. For example, the combination of three-fold screw axis operation $6_3$ and inversion generate mirror reflection $M_z$ with the mirror plane contains Mo triangle, which doesn't contain the inversion center. It is the operator $\mathcal{A} = M_z$ that bunches the four states $\psi$, $PT\psi$, $P\psi$, $T\psi$ together at the four TRI-k points in $k_z = \pi$ plane, corresponding to one cubic DP and three linear DPs.

***Dirac Fermions with different types of band dispersion:*** Fig. 2a illustrates the DFT-calculated band dispersion of $Tl(MoTe)_3$ – a representative of quasi-1D $A(MoX)_3$ compounds that are immune to Peierls distortion. We find that the dispersion within either $k_z = 0$ (Γ-M-K-Γ) or $k_z = \pi$ plane (A-H-L) is relatively flat. In contrast, the bands along the $c$ axis (Γ-A and Γ-L) are dispersive, indicating the quasi-1D feature of the structure. Specifically, there are two steep and linear bands from conduction and valence band meeting at the A(0, 0, 0.5) and L(0.5,0,0.5) point from Γ. Note that within the gap window of ~0.9 eV around the Fermi level, only the Dirac bands show up within the Brillouin zone. Such a clean band structure is expected to be easy to capture by angle-resolved photoemission spectroscopy (ARPES) measurement. The Fermi velocity of the linear Dirac bands along $k_z$ direction is 5.2 x $10^5$ m/s, a value approaching that of graphene, indicating massless Dirac Fermions with high mobility.

Next we discuss the physical properties of the DPs located at the A and L point, respectively. There are four inequivalent DPs at the TRI k-points of the BZ boundary, including one A point, and three L points, as indicated in Fig. 1(f). Here the 4-fold band crossing originates from the four-dimensional irreducible representation of the DP, protected by non-symmophic space group symmetry related to two sublattices. Namely, the system is invariant under some point group operations with respect to a lattice site followed by a partial translation between sublattice sites. Here, the DPs at A and L are protected by the screw axis symmetry $2_1$, and are thus stable under adiabatic transformations that preserve these symmetry operations.

To capture the dispersion physics of the DPs at A and L, we use a four-band $k \cdot p$ model applied at the high-symmetry points A and L and then confirm the results by DFT calculations. In this case, the matrix representation of the inversion symmetry operator $P$ has the form $P = \pm \tau_x$, while the time-reversal symmetry operator $T$ takes the form $T = i\sigma_y K$, where $\tau$ and $\sigma$ is the Pauli matrix working on the orbital and spin subspace, respectively, and $K$ denotes complex conjugation. Taking the four-band basis {$|A,\uparrow\rangle$, $|B,\uparrow\rangle$, $|A,\downarrow\rangle$, $|B,\downarrow\rangle$} with A/B and $\uparrow/\downarrow$ denoting orbital and spin degree of freedom, the low-energy Hamiltonian of the DP with the implement of both $P$ and $T$-invariance condition is written as the form of 4 x 4 matrix [15,38]:

$$H(\mathbf{k}) = \begin{pmatrix} a_1(\mathbf{k})\sigma_z + a_2(\mathbf{k})\sigma_y + a_5(\mathbf{k})\sigma_x & (a_3(\mathbf{k}) - ia_4(\mathbf{k}))\sigma_z \\ (a_3(\mathbf{k}) + ia_4(\mathbf{k}))\sigma_z & -a_1(\mathbf{k})\sigma_z + a_2(\mathbf{k})\sigma_y + a_5(\mathbf{k})\sigma_x \end{pmatrix}, \quad (1)$$

where $a_{0,5}(\mathbf{k})$ are real even functions of $\mathbf{k}$, and $a_{1,2,3,4}(\mathbf{k})$ are real odd functions of $\mathbf{k}$. By further applying specific rotation symmetries, the DPs could be classified according to their leading order of band dispersion based on different band eigenvalues of the rotation operator. Generally, the eigenvalues of $C_n$ rotation are $\alpha = e^{i\pi(2p+1)/n}$, where $p = 0$, 1, ..., $n$-1. If we use the $p$ value to represent the rotation eigenvalue of the four-band basis, we have a $\{p, q, r, s\}$ list with the time-reversal symmetry condition requiring $r = n$-$p$-1 and $s = n$-$q$-1. Furthermore, the invariance of the Hamiltonian under $C_n$ rotation forces that all the $k$-expansion term $k_+^{m_1} k_-^{m_2}$ in $H(k_+, k_-, k_z)$ vanishes if $m_1 - m_2 \neq p - q$ mod $n$ (a more detailed derivation could be found in Ref. [7,15]). Therefore, the leading order of $k$-dispersion at the DP can thus be written as $p$-$q$ mod $n$.

In A(MoX)$_3$ compounds, the little group of the A point is $C_{6h}$ with a $C_6$ rotation axis. Indeed, we find that the band eigenvalues of $C_6$ at the A point are $\{i, -i, -i, i\}$, which means $\{p, q\} = \{4, 1\}$ (the order does not matter). Therefore, the leading order of in-plane $k$-dispersion is 3, indicating cubic Dirac Fermions at A. It describes Fermions having linear dispersion along the $C_6$ axis ($k_z$ direction), while cubic dispersion within $k_x$-$k_y$ plane (the coordinate system is defined in Fig. 1f). This result is confirmed by DFT calculations shown in Fig. 2b, from where it is clear that the band splitting between Dirac bands disperse cubically along both $k_x$ and $k_y$ direction. The cubic Dirac Fermion with isotropic dispersion is famous in nonlinear self-interacting Dirac model in quantum field theory [39], while here we report an anisotropic cubic Dirac Fermion in a real material for the first time. The effective Hamiltonian in the vicinity of the cubic DP is thus written as:

$$H_{Cubic}(\mathbf{k}) \sim \begin{pmatrix} v_x(k_+^3 + k_-^3)\sigma_x + iv_y(k_+^3 - k_-^3)\sigma_y + v_z k_z \sigma_z & 0 \\ 0 & -v_x(k_+^3 + k_-^3)\sigma_x - iv_y(k_+^3 - k_-^3)\sigma_y - v_z k_z \sigma_z \end{pmatrix}, (2)$$

where $k_\pm = k_x \pm ik_y$ with the origin at the DP, and $v_{x,y,z}$ are independent real coefficients. Such a block-diagonal matrix can be decomposed into two Weyl Hamiltonians, with each being characterized by a topological invariant, i.e., a Chern number, which is defined as the number of monopoles of Berry curvature of a closed 2D surface enclosing the Weyl nodes. In A(MoX)$_3$, the DP at A is composed of two opposite cubic Weyl Fermions [7] carrying Chern numbers +3 and -3 joining without annihilation. In other words, such as a DP can be viewed as a being composed of 6 conventional Weyl Fermions with 3 having left-handed and 3 having right-handed chirality. It is known in high energy physics that the Weyl Fermions with the same chirality cannot be degenerate, so such a 6-fold DP is indeed a "new Fermion", caused by symmetries present specifically in crystals.

Next we consider the L point, which has the little group C$_{2h}$ with a C$_2$ rotation axis. Thus, the band eigenvalue of rotation operator is $\pm i$, and $\{p, q\} = \{0, 1\}$. Therefore, the DP at L has linear dispersion along $k_x$ and $k_y$ direction, as confirmed by DFT calculation shown in Fig. 2(b). The type of linear DP at L in A(MoX)$_3$ is identical to that of hypothetical β-cristobalite BiO$_2$ [11] and distorted spinel BaZnSiO$_4$ [40], but is more experimental accessible because of its stability (will discuss later) and the success history of synthetization [23]. In addition, the DP at L in Tl(MoTe)$_3$ is only 7 meV below the Fermi level, which hopefully could be observed by ARPES.

***Dirac Fermions within the conduction bands:*** The crystal symmetry of A(MoX)$_3$ compounds can also host another type of Dirac Fermions, which is induced by the band inversion between conduction and valence bands with an accidental band crossing inside the BZ. Examples include Na$_3$Bi, Cd$_3$As$_2$ that were verified by ARPES measurement [13,14], and ternary honeycomb materials such as BaYBi (Y = Au, Ag and Cu) [41] and metastable allotropes of Ge and Sn [42] predicted by first-principles calculation. However, because the conduction and valence band of A(MoX)$_3$ family only meet at A and L points, such DPs are actually band crossing of two conduction bands or two valence bands at **Λ** points along Γ-A direction. In Tl(MoTe)$_3$ there are two inequivalent DPs having the energy ~ 300 meV above the Fermi level, as marked by the purple circles in Fig. 2a. Interestingly, although they located close to each other in terms of both momentum and energy, the dispersion properties of the two DPs are quite different. In contrast to the band crossings at A and L that are protected by non-symmophic symmetry, the DPs at **Λ** originate from the inversion of bands with different parities. Thus, the inversion operator takes the form P = $\pm\tau_z$, and the low-energy Hamiltonian of the DP deviates from Eq. (1) accordingly [38]. Given the double group representations of $C_6$, the two-fold degenerate $\mathbf{\Lambda}$ bands have three possibilities: $G_7+G_8$, $G_9+G_{10}$, and $G_{11}+G_{12}$, with the eigenvalues of $C_6$ being $e^{\pm 5\pi i/6}$, $e^{\pm \pi i/6}$, $e^{\pm 3\pi i/6}$, respectively. For the DPs, $\mathbf{\Lambda}_1$ is the crossing point of $G_7+G_8$ and $G_9+G_{10}$ bands, and thus have $\{p, q\} = \{2, 0\}$, indicating a quadratic Dirac Fermion with quadratic in-plane dispersion. While the quadratic Weyl points with the Chern numbers $\pm 2$ are theoretically predicted in time-reversal breaking $HgCrSe_4$ [7] and inversion breaking $SrSi_2$ [22], the quadratic Dirac Fermion composed by two quadratic Weyl Fermions with opposite monopole charges has been never reported before. On the other hand, $\mathbf{\Lambda}_2$ is the crossing point of $G_7+G_8$ and $G_{11}+G_{12}$ bands, and thus have $\{p, q\} = \{2, 1\}$, indicating a linear Dirac Fermion. We note that the linear DP at $\mathbf{\Lambda}_2$ distinguishes with the DP at L in two folds: First, DP at $\mathbf{\Lambda}_2$ originates from band crossing due to band inversion and happens in pairs around a TRI k-point (see Fig. 1f); while DP at L is the touching point between conduction and valence band at the boundary of BZ due to non-symmophic symmetry; second, as well as A and $\mathbf{\Lambda}_1$, DP at $\mathbf{\Lambda}_2$ show isotropic dispersion between $k_x$ and $k_y$ direction because the DPs along z axis feels SO(2) symmetry at small in-plane $\mathbf{k}$, while DP at L has anisotropic linear dispersions along $k_x$ and $k_y$ direction, as shown in Fig. 2b. We summarize the physical characters of the different DPs in $Tl(MoTe)_3$ in Table I.

***Peierls distortion and stable structures in quasi-1D $A(MoX)_3$ compounds:*** As noted above, the group of materials $A(MoX)_3$ with reported $P6_3/m$ structure has been made and is laboratory stable. However, the structure determination done so far [23,43] did not shed light on the possibility of possible Peirles distortion. It is natural to expect that the ideal 1D metallic structures are unstable against Peierls distortion, and could have in the undistorted structure either soft phonon modes or higher energy than the distorted phase. Such distortion could destroy the crystal symmetries responsible for the DP and thus open a gap. To predict realistic quasi-1D DSM candidates we thus studied systematically by DFT the phonon spectra (shown in Fig. 6) and thermodynamic stability (shown in Table III and Fig. 3) of the fifteen $A(MoX)_3$ compound related to Peierls distortion. A number of observations can be made: (i) four compounds $Na(MoS)_3$, $Na(MoSe)_3$, $K(MoSe)_3$ and $Rb(MoSe)_3$ have soft phonons for the undistorted structure, indicating dynamical instability. (ii) All six $A(MoX)_3$ compounds with A = Na, K, Rb and X = S and Se, including the four compounds having soft phonon modes, are highly unstable in undistorted structure. As a result, the ground states of these materials experience Peierls transition to lower the total energy by 5.5-6.9 meV/f.u., and thus become semiconducting. (iii) Tl(MoTe)$_3$ and Rb(MoTe)$_3$ are immune to Peierls distortion, leading to the retention of the CDSM phase. (iv) There is an intermediate phase with seven materials. Their undistorted and distorted structures are both dynamically stable and have somewhat similar total energies, implying the coexistence of both phases.

Several A(MoX)$_3$ compounds are not dynamically stable in the high-symmetry P6$_3$/m structure, e.g., K(MoSe)$_3$. The phonon spectrum of undistorted K(MoSe)$_3$ is shown in Fig. 4a. By analyzing the evolution of the phonon eigenvectors we find that one soft phonon mode is found at $\Gamma$ as well as $k_z = 0$ plane, indicating that such high-symmetry metallic structure is dynamically unstable. The eigenvectors of the soft phonon modes of undistorted K(MoSe)3 (see Fig. 4b) show that Peierls distortion naturally happens. The Mo triangles tend to become pairs by moving towards to each other, forming alternative short-long-short-long bonding with each other. Interestingly, the 3 Se atoms within the same plane of each Mo triangle tend to move oppositely and thus forms buckled in-plane structure. We applied such distortion mode to the undistorted structure and after relaxation we find that such Peierls distortion indeed eliminates the negative phonon modes (see Fig. 4c) as well as lowers the total energy by 6.8 meV per f.u.. The distorted structure (see Fig. 1d) has a reduced symmetry with a space group of P$\bar{3}$ (No. 147), in which the screw axis symmetry is no longer preserved. As a result, a band gap is opened at the A and L points. Fig. 4d shows that in distorted K(MoSe)$_3$ there is a 280 meV band gap throughout $k_z = \pi$ plane, indicating that the relative small change in band length between Mo triangles (0.03 Å) induces remarkable effect in electronic structure.

Indeed, the existence of Peierls distortion is basically the competition between band eigenvalues and elastic energy. If the gain of occupied band eigenvalues induced by creating a gap is less than the cost of elastic energy by modulating the atomic positions, Peierls distortion will not happen. The subtlety of whether a quasi-1D system would experience Peierls transition is closely related how "1D" the system is. To demonstrate this, we investigate the relationship between the stability of undistorted A(MoX)$_3$ compounds and their character of one-dimensionality. Since the linear Dirac bands along $\Gamma$-A are mainly contributed by $d_{xz}$ and $d_{yz}$ orbitals of Mo atom, these states strongly extend along the chain direction through the inter-triangle Mo-Mo bonding, leading to large dispersion. On the other hand, the energy bands within a-b plane is rather narrow, implying weak in-plane hopping. Especially, the flatness of the bands within $k_z = \pi$ plane reflects the instability due to Fermi nesting. Therefore, we define a parameter $\lambda = W_c/W_{ab}$ to quantify the anisotropy between c axis and a-b plane, where $W_c$ and $W_{ab}$ denote the widths of the linear Dirac band along $\Gamma$-A and the flat band within $k_z = \pi$ plane, to find out the relationaship between Peierls distortion and the strength of the "1D-ness". More details about $\lambda$ for the 15 A(MoX)$_3$ compounds are shown in Table III. The phase diagram as a function of $\lambda$ is shown in Fig. 3. A clear trend can be found that the compounds having stable DMS phase or competitive DSM phase comparing with Peierls phase are mostly located at the small $\lambda$ area, indicating that they are less 1D-like. On the other hand, the compounds with $\lambda > 100$ are considered more 1D-like, and are thus stabilized as Peierls phase (except Rb(MoTe)$_3$). In addition, for the group of A(MoX)$_3$ with the same A cation, the A(MoSe)$_3$ is the most unstable comparing with X = S and Te.

Finally, we predict that the quasi-1D feature of the A(MoX)$_3$ family provides a platform for realizing low-dimensional physics as well as new electronic and optoelectronic device concepts. Specifically, the optical properties of DSM compounds are uniquely interesting because of the coexistence of reduced dimensionality 1D-metal embedded in a semiconductor bulk, leading to a polarization-dependent optical signature. As shown in Fig. 5a, for Tl(MoTe)$_3$ there is an absorption peak starting from 0 eV for $z$-polarized light originating from the interband optical transitions between the linearly dispersed valence and conduction Dirac bands, while nearly zero absorption for $x$- and $y$-polarized light up to 0.5 eV, originating from the insulating states along these directions. The case of Peierls semiconductor, e.g., K(MoSe)$_3$, is slightly different in that the $z$-direction behaves like a small-gap semiconductor, evidenced by nonzero energy onset of the absorption peak shown in Fig. 5b. Future measurement of the polarization dependence will serve to disentangle the basically different optical properties of 1D-metal and the semiconductor.

**Discussion**

***Metal-insulator transition:*** We expect that Peierls distortion is more noticeable in low temperature, while in high temperature the high-symmetry metallic state is usually more stable because the gain of band eigenvalues is reduced by the thermal excitation of electrons across the band gap. Given that the structural determination of single crystal assigns $A(MoX)_3$ to the high-symmetry space group $P6_3/m$ at high temperature, the message from Fig. 3 could provide some implications on the metal-insulator transition of this system that can be compared with the transport measurements [27,28] and theoretical explanations using electron-phonon coupling [25]. We conclude that the compounds in the "Peierls semiconductor" region should undergo a phase transition from metal to semiconductor when the temperature goes below a critical value, while at low temperature the compounds in the "Coexistence" region could appear both phases based on the growth condition. The phase diagram with three regions shown in Fig. 3 is in great agreement with the measurements made by previous experiments, including that $Tl(MoTe)_3$ is metallic at low temperature; $A(MoSe)_3$ compounds with A = Na, K, Rb undergo a metal-insulator transition below 200 K [28], while $Tl(MoSe)_3$, $In(MoSe)_3$ has at least two types of sample with one metallic down to low temperature and another having phase transition a critical temperature [27]. Our results on thermodynamically stability are also consistent with the recent calculation using the electron-phonon coupling, concluding that the main mechanism of the metal-insulator transition is the dynamic charge density wave that corresponds to Peierls-type displacement [25].

***Experimental accessibility:*** The $A(MoX)_3$ compounds have been synthesized over 30 years ago [23] at 1000-1200 °C in sealed molybdenum crucibles under low argon pressure. The single crystals were needle-like growing along the c axis, indicating the 1D character. Most of this family of compounds were found to be stable in air. It is noticeable that $Na(MoSe)_3$, $In(MoSe)_3$ and $Tl(MoSe)_3$ were synthesized and reported to be a quasi-1D superconductor with relatively high critical fields [27,44,45]. Some compounds from this family were shown to have metal-insulator transition below a critical temperature [25,27,28]. According to our DFT calculations, the compounds that do not have Peierls distortion and thus tend to stay metallic are $Tl(MoTe)_3$ and $Rb(MoTe)_3$. These tellurides are also predicted to have stronger dispersion than selenides and sulfides (see Fig. 7), providing a better chance to resolve the bands by ARPES.

Another challenge for detecting the cubic Dirac fermion is the possibility that the A cation in $A(MoX)_3$ will show some off-stoichiometry, e.g., deficiency which is a natural consequence of the high temperatures required in crystal growth. Indeed, single crystal diffraction showed that while the structures of the quasi-1D MoX chain is nearly perfect, the A-site shows up to 15% under stoichiometric [45]. One might be able to re-plenish the A-site deficiency on the surface by deposition of the cation ultra-high vacuum evaporation so as to make the DP measureable by ARPES. Such a cation deficiency is expected to shift the Fermi energy downwards relative to the stoichiometric crystal, leading to unoccupied Dirac states that are not accessible to ARPES. Fortunately, the cubic DP of $Tl(MoTe)_3$ is calculated to be 226 meV below the Fermi level, so the downward shift may still place the DP near the Fermi energy.

***Remarks on surface states:*** The link between the topological invariant of the bulk and the surface/edge states, known as bulk-boundary correspondence, is the central property of a topological system. For example, in WSMs there are Fermi arcs robustly protected at the surface because of a topological origin [3]: each 2D plane that lies between a pair of Weyl points, perpendicular to the separation between them, is a 2D Chern insulator associated with quantum Hall effect. The edge states of these Chern insulators connect to form surface Fermi arcs, which end at the projection of bulk Weyl points. On the other hand, since the DP in DSM could be understood as two degenerate Weyl points with opposite monopole charge, it has been expected that there are two copies of Fermi arcs on DSM surface forming a ring with two singularities at the surface projection of the DPs in bulk. However, a recent theoretical study reveals that the Fermi arcs on DSM surface are not topologically protected and can be continuously deformed into the case of a topological or normal insulator without any symmetry breaking [46]. Then the surface behavior of a DSM will follow the direction where a small perturbation will lead the system to. For DSMs with a pair of DPs located away from the TRI k-points and $P = \pm \tau_z$ (parity inversion), such as $Na_3Bi$ and $Cd_3As_2$, a small perturbation can open the gap while preserve the band inversion, leading to a topological insulator (TI). Thus, the surface states are robust as a closed Fermi contour.

***Why are cubic Dirac Fermions interesting relative to conventional Dirac Fermions?*** DSMs with DP located at the TRI *k*-points, such as $BiO_2$ (in a hypothetical $SiO_2$ structure [11]) and the A(MoX)$_3$ family, behave as a quantum critical point, rendering it an ideal platform to realize other topological phases by the symmetry tuning [11,15,47]. The main points of interest are: (i) By breaking time-reversal symmetry in DSM, e.g., via introduction of magnetic ions, such systems are known to transform to WSMs. More specifically, for CDSM this was predicted theoretically [15] to lead to WSM with the unique occurrence of three Fermi arcs connecting the surface projection of the Weyl node and its antinode. This is the largest number of pairs of Weyl nodes that can theoretically be accommodated, leading to an enhanced conductivity step for the quantum anomalous Hall effect. (ii) The dispersion power $n$ (= numbers of monopole charges) present in symmetry-broken DSM was predicted to produce $n$-dependent quantum interference effects [48], leading to dispersion-dependent quantum transport phenomena. For conventional DSM/WSM ($n = 1$), a destructive quantum interference results in a weak antilocalization correction proportional to $-\sqrt{B}$ in the weak field limit. Such negative longitudinal magnetoresistance, also known as chiral anomaly [49,50], has been confirmed by various transport measurements [51-55]. In contrast, for DSM/WSM with $n$ = 2, it has been predicted that a weak localization correction proportional to $+\sqrt{B}$ applies on the magnetoconductivity [48], calling for experimental verifications. With the material realization of CDSM ($n = 3$), its transport behavior becomes an accessible and open question. (iii) The stronger screening of bare interaction and disorder in DSM/WSM with high-order dispersions provides the opportunity for more exotic physics, such as quantum criticality and phase transition [56-58]. In DSM/WSM, the density of states at the band crossing point behaves as $\rho(E) \sim |E|^{2/n}$. Compared with the linear-dispersing direction, the quadratic and cubic-dispersing directions have enhanced density of states near the band crossing point, which results in stronger screening. Specifically, in CDSM ($n = 3$) the Coulomb interactions along the in-plane directions are screened with a faster decaying than that along the rotation axis ($r^{-1}$). Recently, it was predicted that WSM with $n = 3$, in the presence of short-range interactions, can easily undergo a continuous quantum phase transition into either a translational symmetry breaking axion insulator or a rotational symmetry breaking nematic state [59]. Furthermore, the nonlinear dispersion and the 1D nature of a condensed matter system would cause a breakdown of the interacting Fermi liquid theory for electron behavior, leading to Luttinger liquid instead.

**Acknowledgements**

We thank L. Fu, V. Juricic, J. Liu, X. Zhang, M. Hoesch and X. Zhou for helpful discussions. This work was supported by the U.S. Department of Energy, Office of Science, Basic Energy Sciences, Materials Sciences and Engineering Division under Grant No. DE-FG02-13ER46959 to University of Colorado, Boulder. This work used resources of the National Energy Research Scientific Computing Center, which is supported by the Office of Science of the U.S. Department of Energy under Contract No. DE-AC02-05CH11231.

Table I: Different types of Dirac points and their physical properties in A(MoX)$_3$ compounds. Note that the energy of DP is related to the Fermi level, and $w = e^{i\pi/6}$.

| Wave vector of DP | Multiplicity in BZ | Energy (eV) | In-plane dispersion | Little group of DP | Inversion operator P | Rotation eigenvalue |
|---|---|---|---|---|---|---|
| A(0,0,0.5) | 1 | -0.226 | Cubic | C$_{6h}$ | $\pm\sigma_x$ | $\{i, -i, -i, i\}$ |
| L(0.5,0,0.5) | 3 | 0.007 | Linear | C$_{2h}$ | $\pm\sigma_x$ | $\{i, -i, -i, i\}$ |
| $\boldsymbol{\Lambda}_1$(0,0,0.154) | 2 | 0.294 | Quadratic | C$_6$ | $\pm\sigma_z$ | $\{w, w^5, w^{-1}, w^{-5}\}$ |
| $\boldsymbol{\Lambda}_2$(0,0,0.136) | 2 | 0.328 | Linear | C$_6$ | $\pm\sigma_z$ | $\{w, w^{-1}, i, -i\}$ |

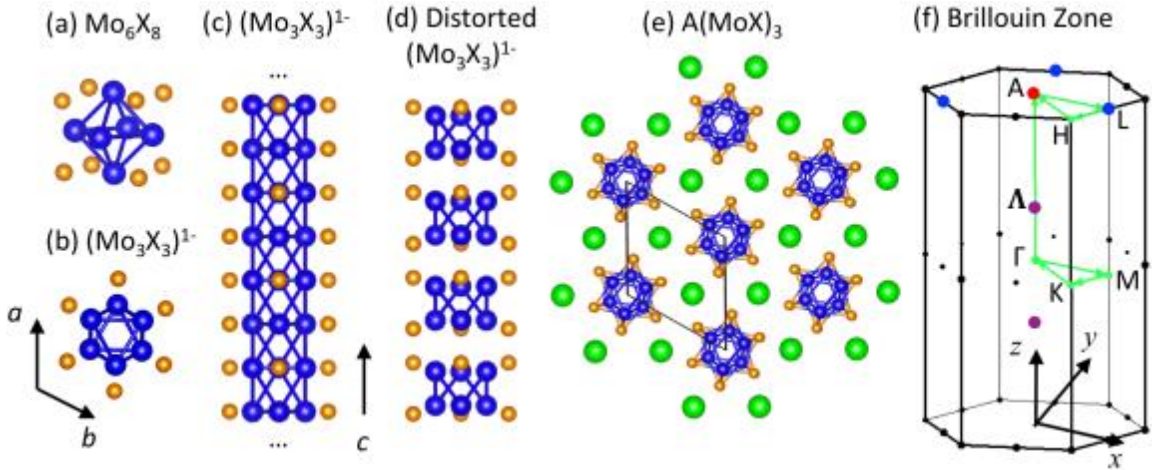

Fig. 1 (a) Structure of the ($Mo_6X_8$) cluster unit with a $Mo_6$ octahedral surrounded by a cubic $X_8$ cage. (b) Top and (c) side view of the quasi-1D chain ($Mo_3X_3$)$^{1-}$. (d) Side view of the quasi-1D chain ($Mo_3X_3$)$^{1-}$ with Peierls distortion. (e) Crystal structure of the cluster compound A(MoX)$_3$. The blue, orange and green balls denote Mo, X and A atom, respectively. (f) Hexagonal Brillouin zone and the high-symmetry *k*-path for band structure calculation. The dots mark the inequivalent Dirac points (one A point, three L points and two **Λ** points representing **Λ**$_1$ and **Λ**$_2$) in undistorted structure.

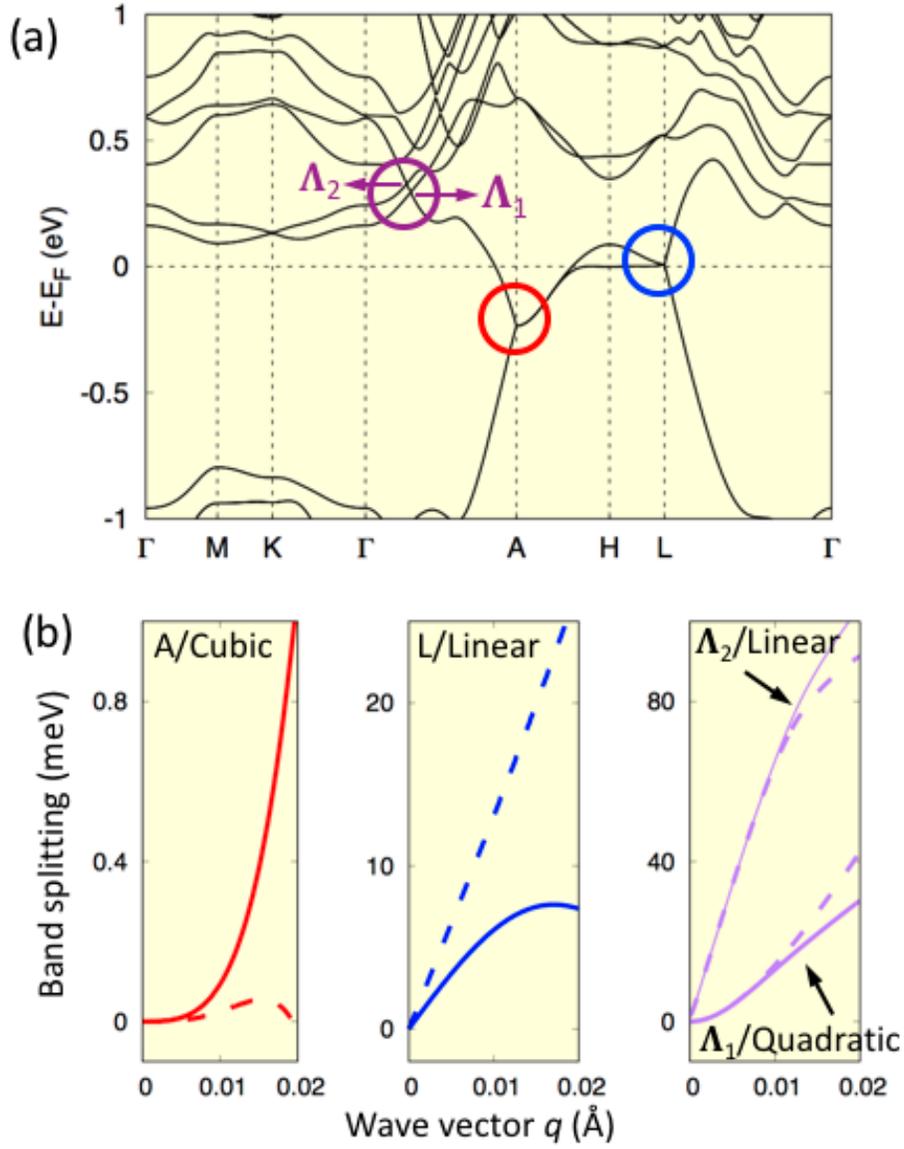

Fig. 2 (a) DFT Calculated band structure of Tl(MoTe)₃ as a representative of stable A(MoX)₃ compounds. The different types of Dirac point are marked by red, blue and purple circles. (b) In-plane band splitting in the vicinity of the four types of Dirac Fermions at A, L, $\Lambda_1$ and $\Lambda_2$ point. The solid (dash) lines denote DFT band splitting along $k_x$ ($k_y$) direction, within the coordinate system defined in Fig. 1f.

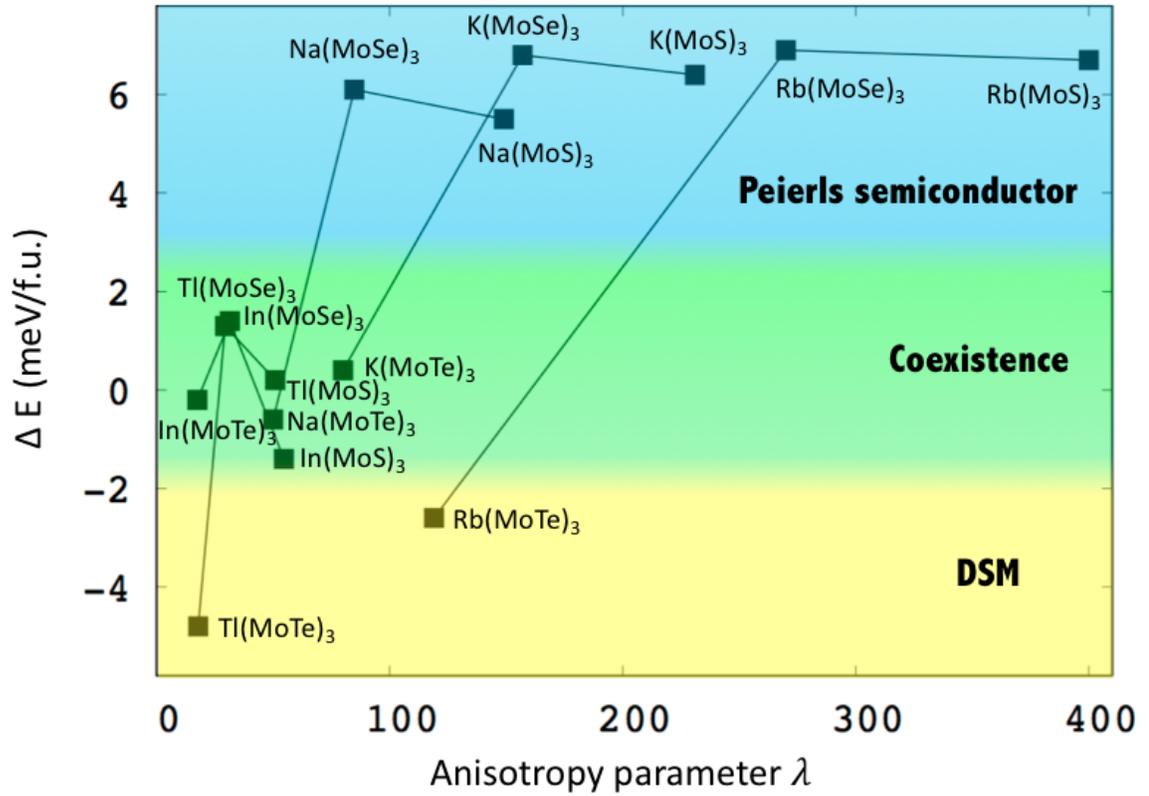

Fig. 3 (a) Phase diagram of DSM vs Peierls distortion as a function of anisotropy parameter $\lambda$ for 15 A(MoX)$_3$ compounds. The black lines connecting compounds with the same A cation are the guide to eyes.

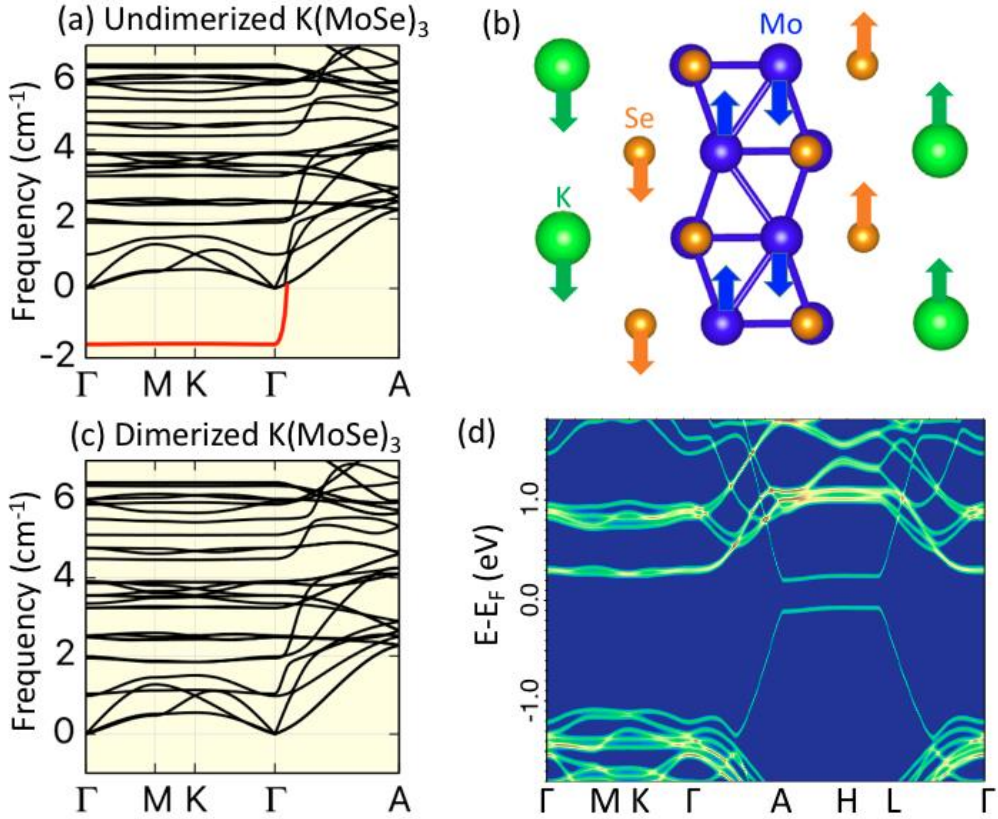

Fig. 4 Calculated phonon spectrum of undistorted K(MoSe)$_3$. The red phonon dispersion indicates soft phonon modes. (b) Vibration modes of the soft phonon at the $\Gamma$ point in (a). (c) Calculated phonon spectrum of K(MoSe)$_3$ under Peierls distortion with the soft mode eliminated. (d) Band dispersion of distorted K(MoSe)$_3$ shows a band gap throughout $k_z = \pi$ plane (A-H-L).

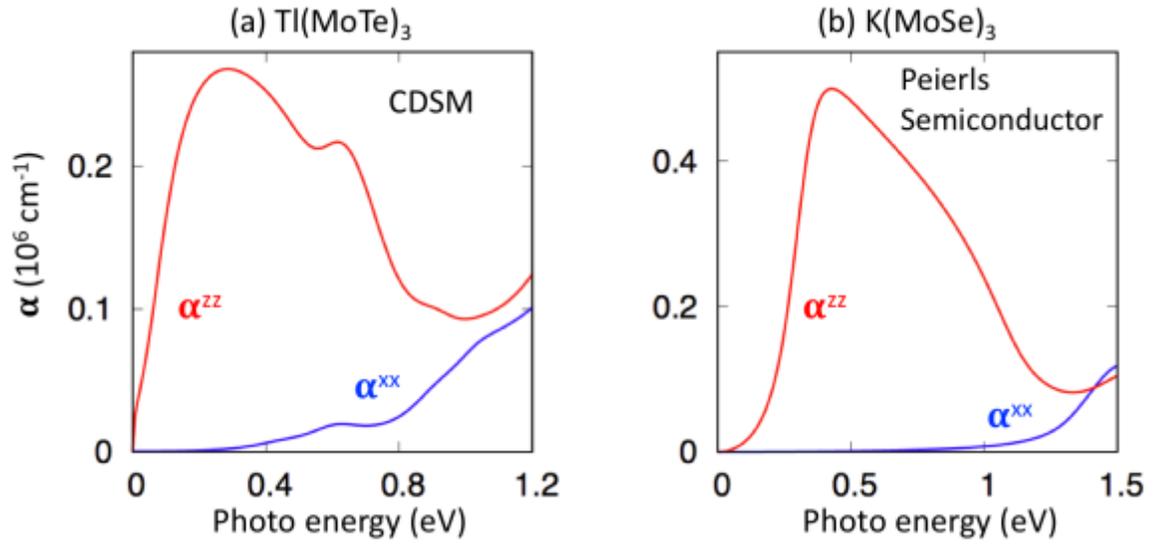

Fig. 5 Absorption coefficient ($\alpha^{xx} = \alpha^{yy}$; $\alpha^{zz}$) of (a) Tl(MoTe)$_3$ and (b) K(MoSe)$_3$ from DFT calculations.

# Appendix

**A. Classification of different types of Fermions in solid state physics**

Table II shows the examples of materials that host different types of Fermions classified by the degree of degeneracy ($g$) and the highest power of band dispersion ($n$). Here we consider three-dimensional crystals with spin-orbit coupling, respecting time-reversal symmetry. We consider single-point degeneracy in the $k$-space, so the materials with line node is not included. Some cases, e.g., $g = 8$ and $n = 3$, are forbidden because of the restriction of crystal symmetries. Some cases are predicted to exist but no material realization yet, marked by "?" in the table. The materials with asterisk have hypothetical configurations, while all of the rest examples (including our work) have been synthesized as single crystals.

Table II: Different types of Fermions in solid state physics with example materials.

| $g$ / $n$ | 2 (e.g. WSM) | 3 | 4 (e.g. DSM) | 6 | 8 |
|---|---|---|---|---|---|
| 1 | TaAs [9], WTe$_2$ [8] | Pd$_3$Bi$_2$S$_2$ [5], WC [19] | Na$_3$Bi [4], BiO$_2$* [11] | MgPt [5], Li$_2$Pd$_3$B [5] | Ta$_3$Sb [5], Bi$_2$AuO$_5$ [18] |
| 2 | SrSi$_2$ [22] | ZrTe [19], TaN [17] | $\alpha$-Sn [65] | PdSb$_2$ [5] | Forbidden |
| 3 | ? | Forbidden | A(MoX)$_3$ (this work) | Forbidden | Forbidden |

**B. First-principles calculations**

All The calculations including total energy, electronic structure and phonon dispersion were performed by density functional theory (DFT) where the geometrical and total energies are calculated by the projector-augmented wave (PAW) pseudopotential [60] and the exchange correlation is described by the generalized gradient approximation of Perdew, Burke and Ernzerhof (PBE) [61] as implemented in the Vienna ab initio package (VASP) [62]. The plane wave energy cutoff is set to 500 eV, and the electronic energy minimization was performed with a tolerance of $10^{-5}$ eV. Spin-orbit coupling is taken into account self-consistently throughout the electronic structure calculations. The atomic projection on band structure is calculated by projecting the wave functions with plan wave expansion on the orbital basis (spherical harmonics) of each atomic site. The phonon spectra were identified by PHONOPY package [63], in which the force constants are calculated in the framework of density-functional perturbation theory (DFPT) [64]. Phonon calculations were performed within a 1×1×4 supercell (56 atoms). To evaluate the anisotropic optical properties of quasi-1D A(MoX)$_3$ compounds, we calculate the frequency ($\omega$) dependent dielectric function $\varepsilon^{ij}(\omega)$ based on DFT. Then the optical absorption coefficient is evaluated from the dielectric function:

$$\alpha^{ij}(\omega) = \frac{\sqrt{2}\omega}{c}\sqrt{|\varepsilon^{ij}(\omega)| - Re[\varepsilon^{ij}(\omega)]}, \tag{3}$$

where $c$ is the speed of light and $Re[\varepsilon^{ij}(\omega)]$ is the real part of $\varepsilon^{ij}(\omega)$.

In Table III, we show the geometry, electronic and stability signatures of undistorted and distorted A(MoX)$_3$ structures by DFT calculations. The results indicate that 9 out of the 15 compounds have comparative total energies for the two structures taken into consideration, and they are in general more 3D-like, as suggested by the anisotropy parameter $\lambda$.

Fig. 6 shows the phonon spectra of 15 A(MoX)$_3$ compounds with undistorted P6$_3$/m structure. We find that four compounds Na(MoS)$_3$, Na(MoSe)$_3$, K(MoSe)$_3$ and Rb(MoSe)$_3$ are dynamically unstable because of the negative phonon modes for the $k_z = 0$ plane of the undistorted P6$_3$/m structure.

Fig. 7 shows the band structure of 15 A(MoX)$_3$ compounds with undistorted P6$_3$/m structure. We find that all the quasi-1D compounds have strong dispersions along the $c$ axis and flat in-plane dispersions. Comparatively, the compounds with heavier A and X elements tend to have stronger in-plane dispersions because of the relativistic effects.

Table III: Calculated inter-triangle Mo-Mo bond length ($d_{int}$) of undistorted and distorted structures, $z$-direction Fermi velocity $v_F$ at the A point for undistorted structure, anisotropy parameter $\lambda$ for undistorted structure and Peierls stabilization energy $\Delta E$ (the total energy difference between undistorted and distorted structures) for 15 A(MoX)$_3$ compounds.

| Compounds | Undistorted $d_{int}$ (Å) | Distorted $d_{int}$ (Å) | $v_F$ (m/s) | $\lambda$ | $\Delta E$ (mev/f.u.) |
|---|---|---|---|---|---|
| Na(MoS)$_3$[*] | 2.690 | 2.666/2.714 | 8.1 x 10$^5$ | 149 | 5.5 |
| Na(MoSe)$_3$[*] | 2.719 | 2.691/2.745 | 8.3 x 10$^5$ | 84.8 | 6.1 |
| Na(MoTe)$_3$ | 2.762 | 2.730/2.801 | 7.0 x 10$^5$ | 50 | -0.6 |
| K(MoS)$_3$ | 2.694 | 2.669/2.719 | 8.4 x 10$^5$ | 231 | 6.4 |
| K(MoSe)$_3$[*] | 2.723 | 2.692/2.753 | 8.0 x 10$^5$ | 157 | 6.8 |
| K(MoTe)$_3$ | 2.764 | 2.749/2.781 | 7.3 x 10$^5$ | 80 | 0.4 |
| Rb(MoS)$_3$ | 2.696 | 2.672/2.720 | 9.0 x 10$^5$ | 400 | 6.7 |
| Rb(MoSe)$_3$[*] | 2.724 | 2.696/2.753 | 8.3 x 10$^5$ | 270 | 6.9 |
| Rb(MoTe)$_3$ | 2.765 | 2.763/2.768 | 7.3 x 10$^5$ | 119 | -2.6 |
| In(MoS)$_3$ | 2.702 | 2.678/2.725 | 8.2 x 10$^5$ | 54.6 | -1.4 |
| In(MoSe)$_3$ | 2.724 | 2.701/2.746 | 6.9 x 10$^5$ | 31.4 | 1.3 |
| In(MoTe)$_3$ | 2.759 | 2.736/2.782 | 5.4 x 10$^5$ | 17.5 | -0.2 |
| Tl(MoS)$_3$ | 2.699 | 2.676/2.722 | 8.0 x 10$^5$ | 50.9 | 0.2 |
| Tl(MoSe)$_3$ | 2.723 | 2.701/2.746 | 6.7 x 10$^5$ | 30.7 | 1.34 |
| Tl(MoTe)$_3$ | 2.760 | 2.735/2.785 | 5.2 x 10$^5$ | 17.8 | -4.8 |

[*]Undistorted structure is dynamically unstable, i.e., has soft phonon modes.

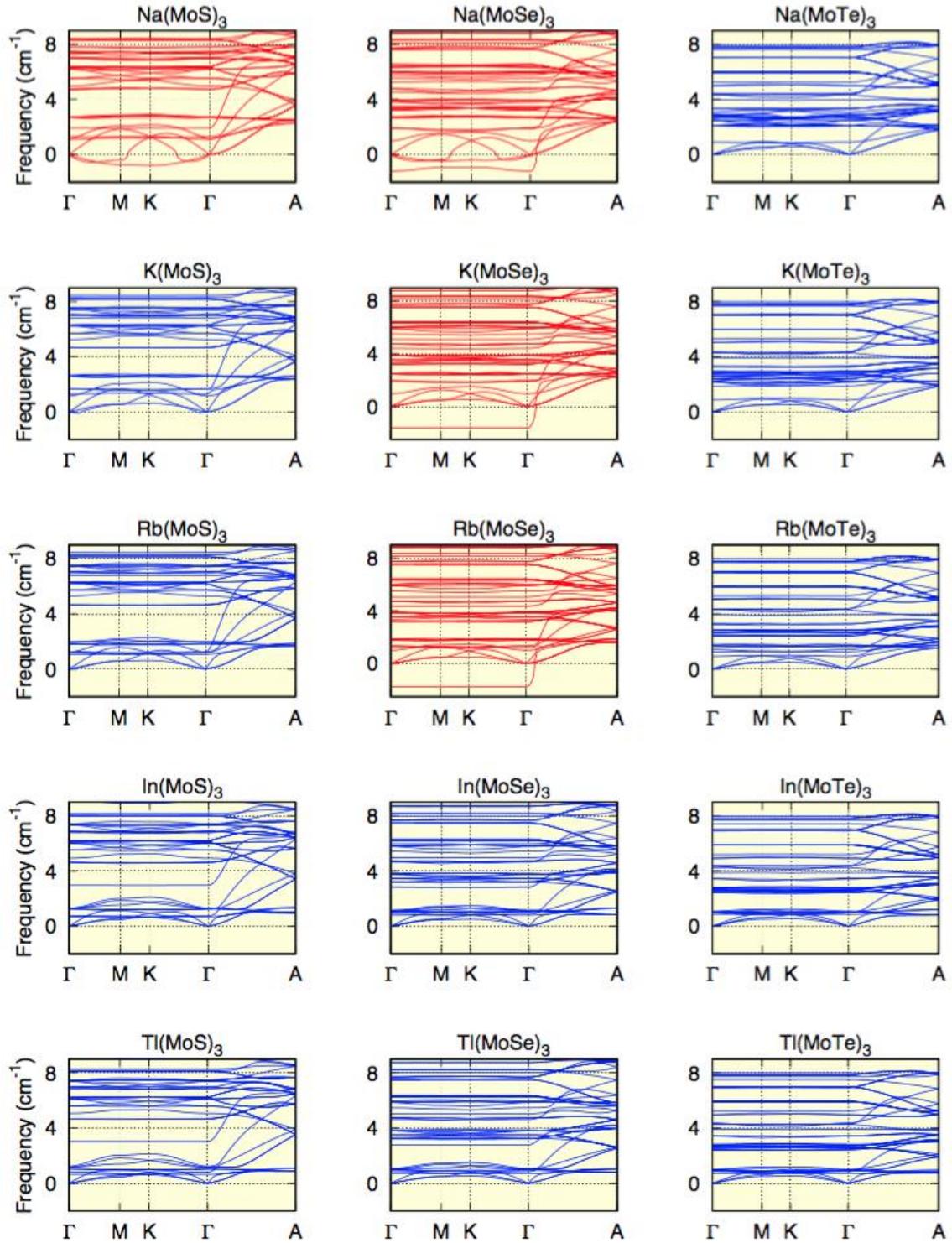

Fig. 6: Phonon spectra of 15 A(MoX)₃ compounds with undistorted P6₃/m structure. The red and blue dispersion indicate spectra with soft phonons and without soft phonons, respectively.

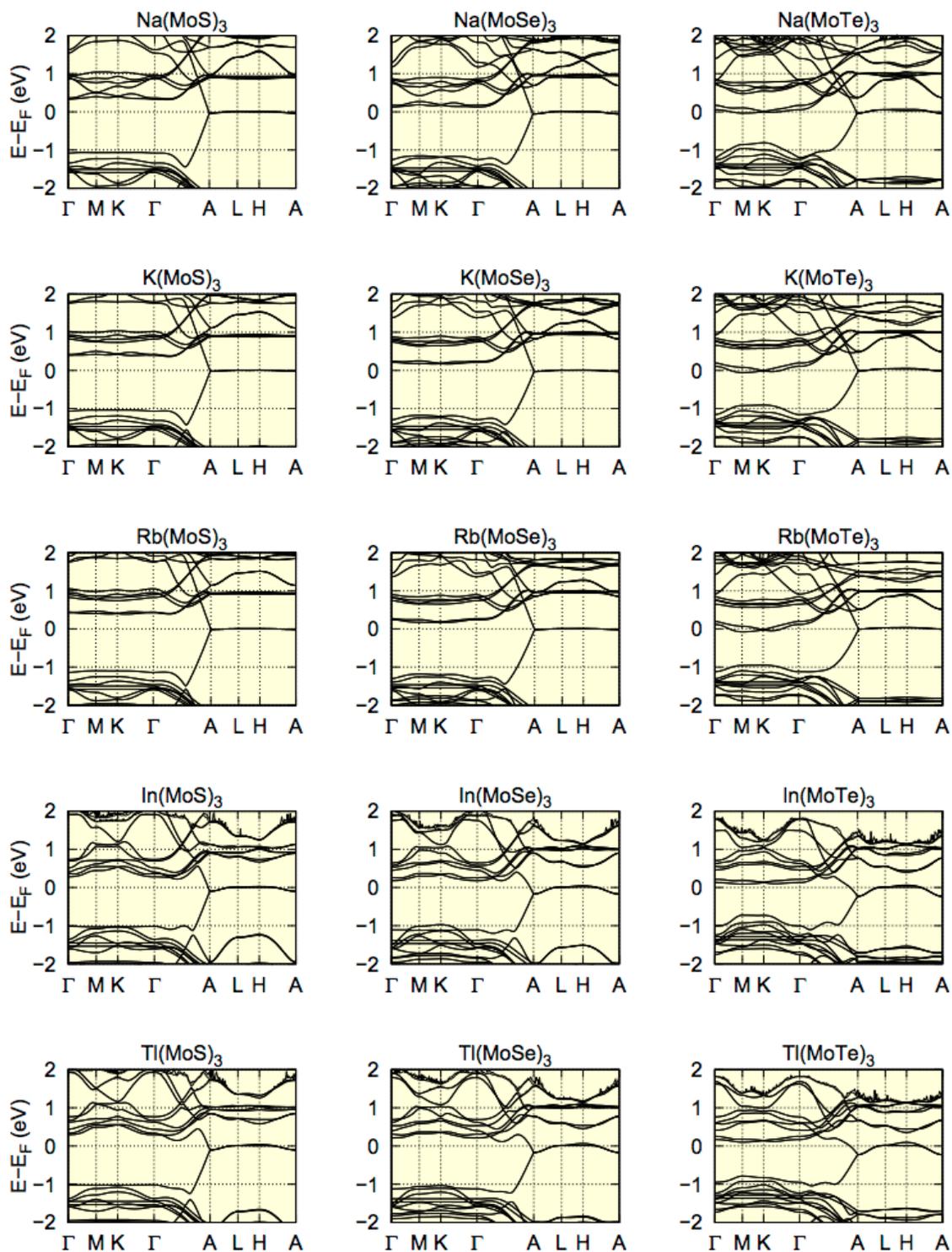

Fig. 7: Band structure of 15 A(MoX)$_3$ compounds with undistorted P6$_3$/m structure.

**C. Space groups that host cubic Dirac fermions**

Here we use symmetry analysis to show that how non-symmophic symmetry ensures four-fold degenerate, i.e., Dirac point (DP), at certain time-reversal invariant (TRI) $k$-points in a spin-orbit system preserving both inversion symmetry $P$ and time-reversal symmetry $T$. Then we show that out of 230 space groups only P6$_3$/m (No. 176) and P6/mcc (No. 192) have appropriate symmetries to host cubic Dirac fermions.

In a spin-orbit system, the anti-unitary operator $T$ behaves as $T^2 = -1$, leading to Kramers degeneracy. Together with inversion symmetry, it turns out that all the energy bands are two-fold degenerate with the two components related each other by $PT$, i.e., $\psi(\boldsymbol{k}, \boldsymbol{\sigma})$ and $PT\psi(\boldsymbol{k}, \boldsymbol{\sigma}) = \psi(\boldsymbol{k}, -\boldsymbol{\sigma})$, known as spin degeneracy. Therefore, to achieve four-fold degeneracy we need an extra pair of state $L\psi$ and $PTL\psi$ with $[L, H] = 0$ that differ with $\psi$ and $PT\psi$, while $L$ is a Hermitian symmetry operator of the system. We are thus looking for another Hermitian symmetry operator $\mathcal{A}$ to fulfill that $\{A, A_{PT}\} \cap \{A_L, A_{LPT}\} = \emptyset$, where $A$ is the eigenvalue of $\psi$ under $\mathcal{A}$. By achieving this we get two pairs to bands $\{\psi, PT\psi\}$ and $\{L\psi, PTL\psi\}$ that have different eigenvalues of $\mathcal{A}$, so they must have a band crossing rather than opening a gap. The task is basically to find the two operators $L$ and $\mathcal{A}$, and the degeneracy will happen at the k-points invariant with these two symmetry operations. For example, if the $k$-points that are invariant under both L and $\mathcal{A}$ forms a line, the system is thus a nodal-line semimetal.

Without adding new symmetries, we first let $L = P$. Since $P$ operator reverses the momentum, there are only eight TRI $k$-points in the BZ are $P$-invariant. Now we are looking for operator $\mathcal{A}$ that fulfills

$$\{A, A_{PT}\} \cap \{A_P, A_T\} = \emptyset. \tag{C1}$$

We next consider the most common two-fold symmetries for $\mathcal{A}$ that all the TRI $k$-points can preserve, which have two eigenvalues. From Eq. (C1) we have $A = -A_P$, indicating

$$\mathcal{A}P\psi = A_P P\psi = -PA\psi = -P\mathcal{A}\psi, \tag{C2}$$

which leads to anti-commutation relationship

$$\{\mathcal{A}, P\} = 0. \tag{C3}$$

Given that $P$ is commute with any point group operations, we conclude that $\mathcal{A}$ contains a non-symmophic symmetry that is a combination of point group operation and fractional translation. In addition, from Eq. (1) there is another condition $A = -A_T$. Considering $[\mathcal{A}, T] = 0$, we have

$$\mathcal{A}T\psi = T\mathcal{A}\psi = TA\psi = -TA_T\psi = A_T T\psi, \tag{C4}$$

which indicates $A_T = \pm i$ and thus

$$\mathcal{A}^2 = -1. \tag{C5}$$

Therefore, the symmetry operation $\mathcal{A}$ that fulfills Eq. (C3) and (C5) ensures a DP in certain TRI $k$-points.

Combining three symmetry filters for cubic Dirac semimetal, i.e., inversion, $C_6$ and non-symmophic symmetry, only four possibilities, P6$_3$/m (No. 176), P6/mcc (No. 192), P6$_3$/mcm (No. 193) and P6$_3$/mmc (No. 194) are left. All of these space groups have DPs at the four TRI k-points (one A point and three L points) within $k_z = \pi$ plane. For space groups No. 176, 193 and 194, there is an axis symmetry {C$_2$|(0,0,1/2)}, which transform (x, y, z) in position space to (-x, -y, z+1/2). Considering the combination symmetry $\mathcal{A}$ = P{C$_2$|(0,0,1/2)}, it is easy to test that $[\mathcal{A}, P] = 0$ at $k_z = 0$ plane and {$\mathcal{A}$, P} = 0 at $k_z = \pi$ plane. On the other hand, $\mathcal{A}^2$ preserves (x, y, z) while rotates spin by $2\pi$, leading to a minus sign that $\mathcal{A}^2 = -1$. Therefore, $\mathcal{A}$ protects the four-fold degeneracy at the four TRI k-points within $k_z = \pi$ plane. However, space groups No. 193 and 194 have three mirror planes parallel to C$_6$ axis, posing extra symmetry conditions that force three high-symmetry lines to be degenerate. Here we still take $\mathcal{A}$ = P{C$_2$|(0,0,1/2)} but L = $M_x$, which transform (x, y, z) in position space to (-x, y, z). The commutation relationship then reads {$\mathcal{A}$, $M_x$} = 0 and [$\mathcal{A}$, PT$M_x$] = 0 at $k_z = \pi$ plane. In this case $\mathcal{A}$ and L keep the whole $k_x = 0$ line as well as another two lines related by C$_3$ symmetry at $k_z = \pi$ plane, rendering the system as a nodal-line or nodal-ring semimetal.

On the other hand, space group P6/mcc (No. 192) has six glide reflection planes that all contains the C$_6$ axis, and here we take {$M_x$|(0,0,1/2)} that transforms (x, y, z) to (-x, y, z+1/2). Similarly, considering the combination symmetry $\mathcal{A}$ = $P${$M_x$|(0,0,1/2)}, we also have {$\mathcal{A}$, $P$} = 0 at $k_z = \pi$ plane and $\mathcal{A}^2 = -1$ that protects only four DPs, and no extra symmetries for more degenerate k-points. Finally, we reach the conclusion that out of 230 space groups only P6$_3$/m (No. 176) and P6/mcc (No. 192) have appropriate symmetries to host cubic Dirac fermions.